\documentstyle[12pt,amssymb]{article}

\newcommand{\eq}[1]{(\ref{#1})}

\def\e{{\rm e}}

\def\l{\left(}
\def\r{\right)}

\def\zd{\zeta_d}
\def\zu{\zeta_u}
\def\ep{\e^{i\phi}}
\def\emi{\e^{-i\phi}}

\begin{document}
\title{%
\begin{flushright}
\normalsize 
ULB-TH-00-29\\
UNIL-IPT-00-27
\end{flushright}
\protect\vspace{5mm}
Three generations on a local vortex in extra
dimensions. 
}
\author{J.-M.Fr\`ere$^1$, M.V.Libanov$^{2,1}$ and S.V.Troitsky$^{2,3,1}$\\
\small\em
$^{1}$~Service de Physique Th\'{e}orique, CP 225,\\
\small\em
  Universit\'{e} Libre de Bruxelles, B--1050, Brussels, Belgium;\\
\small\em
$^{2}$~Institute for Nuclear Research of the Russian Academy of Sciences,\\
\small\em
60th October Anniversary Prospect 7a, 117312, Moscow, Russia;\\
\small\em
$^{3}$~Institute of Theoretical Physics, University of Lausanne,\\
\small\em
CH-1015, Lausanne, Switzerland
\small\em
}
\date{}
\maketitle
\vspace{-12mm}
\begin{abstract}
We develop an approach to the origin of three generations of the
Standard Model fermions from one generation in a higher-dimensional
theory, where four-dimensional fermions appear as zero modes trapped in the
core of a topological defect, and the hierarchy of masses and mixings
is produced by wave function overlaps in extra dimensions. We present
a model with unbroken $U(1)$ symmetry where three zero modes appear on
an Abrikosov-Nielsen-Olesen vortex due to nontrivial scalar--fermion
interactions.
\end{abstract}
\newpage
{\bf 1.} Models of particle physics in more than four spacetime
dimensions, where the Standard Model fields are localized on a
four--dimensional topological defect, or ``brane'', open new
possibilities to explain the misterious pattern of fermion mass
hierarchies, as well as the origin of three generations of fermions
with identical quantum numbers. If different fermionic modes have
different wave function profiles in extra dimensions, then their
overlaps with the Higgs wave function may produce hierarchical
structure of masses and mixings \cite{overlaps}. Of particular
interest are the class of models \cite{LT} where each
multi--dimensional fermion develops three chiral zero modes localized
on a four--dimensional brane. This occurs due to certain topological
properties of the brane background.  The Index theorem guarantees that
the three zero modes are linearly independent, and thus have different
profiles in extra dimensions. Analysis of the equations for these zero
modes demonstrates that a hierarchy in the mass matrix indeed appears
due to overlaps of the wave functions. For the discussion of this
mechanism and
comparison with other approaches, see Ref.~\cite{LT}.

The main drawback of the models of Ref.~\cite{LT} is the complications
needed to generate nonzero inter-generation mixings. In a particular
model, it was necessary to break explicitly a $U(1)$ symmetry, which
gave rise to nontrivial topology of the vortex (which played the role
of a brane). With unbroken $U(1)$, the mixings were zero. Explicitly
broken $U(1)$ could not be gauge symmetry, thus the vortex had to be
global and had to have logarithmically divergent energy.  Furthermore,
the question whether the vortex is stable with broken $U(1)$ remained
open.

In this Letter, we discuss an improved version of the models of
Ref.~\cite{LT}. We present a model, in which a similar mechanism gives
rise to three chiral fermionic generations on a vortex embedded in a
six--dimensional spacetime, with hierarchical pattern of masses and
mixing angles, but the corresponding $U(1)$ symmetry is unbroken. In
this way we overcome the problems mentioned above. The cost for
unbroken $U(1)$ is the necessity to invoke operators of higher
dimensionality in the scalar--fermion interactions. We include,
however, all lowest dimensionality operators consistent with the gauge
symmetries of the Standard Model and the gauge $U(1)$ symmetry which
gives rise to nontrivial topology of the vortex. The requirement of
renormalizability of the theory does not make much sense anyway in the
six-dimensional models, since even usual Yukawa scalar-fermion-fermion
coupling is non--renormalizable. Thus, any multidimensional field
theory is usually considered as a low energy limit of some more
fundamental model which includes gravity as well.

{\bf 2.} 
The matter field content of the model is summarized in Table
\ref{table:fields}. The scalar field $\Phi$, together with $U(1)_g$
gauge field,  forms a vortex, while two
other scalars, $X$ and $H$, develop profiles localized on the vortex.
There is one fermionic generation which consists of five
six--dimensional fermions $Q$, $U$, $D$, $L$, and $E$. Each of the
fermions develop, in the vortex background, three chiral zero modes
localized in the core of the vortex, which correspond to three
generations of the Standard Model fermions. We do not discuss here
subtle issues of localizing gauge fields on the four--dimensional
brane. 

\begin{table}
\begin{center}
\begin{tabular}{|rc|c|c|c|c|c|}
\hline
\multicolumn{2}{|c|}{fields}
& profiles&\multicolumn{2}{|c|}{charges}&
\multicolumn{2}{|c|}{representations}\\
\cline{4-7}
&&&$U(1)_g$&$U(1)_Y$&$SU(2)_W$&$SU(3)_C$\\
\hline
scalar&$\Phi$&$F(r)\e^{i\theta}$&+1&0&{\bf 1}&{\bf 1}\\
&&$F(0)=0$, $F(\infty)=v_\Phi$&&&&\\
\hline
scalar&$X$&$X(r)$&+1&0&{\bf 1}&{\bf 1}\\
&&$X(0)=v_X$, $X(\infty)=0$&&&&\\
\hline
scalar&$H$&$H(r)$&$-1$&$+1/2$&{\bf 2}&{\bf 1}\\
&&$H(0)=v_H$, $H(\infty)=0$&&&&\\
\hline
fermion&$Q$&3 L zero modes&axial $+3/2$&$+1/6$&{\bf 2}&{\bf 3}\\
\hline
fermion&$U$&3 R zero modes&axial $-3/2$&$+2/3$&{\bf 1}&{\bf 3}\\
\hline
fermion&$D$&3 R zero modes&axial $-3/2$&$-1/3$&{\bf 1}&{\bf 3}\\
\hline
fermion&$L$&3 L zero modes&axial $+3/2$&$-1/2$&{\bf 2}&{\bf 1}\\
\hline
fermion&$E$&3 R zero modes&axial $-3/2$&$-1$&{\bf 1}&{\bf 1}\\
\hline
\end{tabular}
\end{center}
\caption{Scalars and fermions with their gauge quantum numbers. For
convenience, we describe here also the profiles of the classical
scalar fields and fermionic wave functions in extra dimensions.}
\label{table:fields}
\end{table}

We use the notations of Ref.~\cite{LT}; in particular, we denote
four--dimensional indices by Greek letters, $x_\mu$, $\mu=0,\dots,3$, and
introduce polar coordinates  $r$, $\theta$ in the $x_4$, $x_5$ plane.

Let us consider the following scalar potential:
$$
V_s=
{\lambda\over 2}\l |\Phi|^2-v^2\r^2+
{\kappa\over 2} \l |H|^2-\mu^2\r^2+h^2|H|^2|\Phi|^2+
{\rho\over 2} \l |X|^2-v_1^2\r^2+\eta^2|X|^2|\Phi|^2.
$$
We study solutions to the classical field equations for the system
involving three scalar fields $\Phi$, $X$ and $H$ as well as 
the $U(1)_g$ gauge field.
Let us restrict ourselves to topologically nontrivial
solutions to the classical field equations which do not depend on $x_\mu$.
For a certain range of parameters, the lowest
energy solution of this kind is given by a vortex configuration build
up of the gauge field and $\Phi$, where
$$
\Phi=\e^{i\theta}F(r).
$$
Two other scalar
fields do not depend on $\theta$, are nonzero at $r=0$ and satisfy the following boundary
conditions:
$$
\left.{dX\over dr}\right|_{r=0}=\left.{dH\over dr}\right|_{r=0}=0, 
~~~~ X(r=\infty)=H(r=\infty)=0.
$$
This solution describes a superconducting bosonic string \cite{Witten}
with currents of two different fields, $H$ and $X$. Far from the core
of the string, $H(r)$ and $X(r)$ fall off exponentially, so the two
scalar fields are localized close to the four-dimensional hypersurface $r=0$.

For the scalar--fermion interactions, we take the most general
operators of the lowest order, consistent with gauge invariance,
$$
V_{sf}=V_1+V_2+V_3,
$$
\begin{equation}
\begin{array}{c}
\displaystyle
V_1=g_q\Phi^3\bar Q{1-\Gamma_7\over 2} Q+
g_u\Phi^{*3}\bar U{1-\Gamma_7\over2} U +
g_d\Phi^{*3}\bar D{1-\Gamma_7\over 2} D+\\
\displaystyle
g_l\Phi^3\bar L{1-\Gamma_7\over 2} L+
g_e\Phi^{*3}\bar E{1-\Gamma_7\over2} E+{\rm h.c.},
\end{array}
\label{VPhiF}
\end{equation}
\begin{equation}
V_2=
Y_dHX\bar Q\frac{1-\Gamma_7}{2}D+
Y_u\tilde HX^*\bar Q\frac{1-\Gamma_7}{2}U+
Y_lHX\bar L\frac{1-\Gamma_7}{2}E+
{\rm h.c.},
\label{V2}
\end{equation}
\begin{equation}
V_3=
Y_d\epsilon_dH\Phi\bar Q\frac{1-\Gamma_7}{2}D+
Y_u\epsilon_u\tilde H\Phi^*\bar Q\frac{1-\Gamma_7}{2}U+
Y_l\epsilon_lH\Phi\bar L\frac{1-\Gamma_7}{2}E+
{\rm h.c.},
\label{V3}
\end{equation}
where $\tilde H_i=\epsilon_{ij}H^*_j$, $i$, $j$ are $SU(2)_W$
indices (we denote Yukawa coupling constants in Eq.~\eq{V2} as
$Y_{u,d,l}$, and in Eq.~\eq{V3} as $Y_u\epsilon_u, \dots$, for
convenience).  

Note that the scalar background $\Phi^3$, where the field $\Phi$ has
the winding number one, $\Phi=F(r)\e^{i\theta}$, has exactly the same
topological properties as the background $\Phi_1$ of the vortex with
winding number three, $\Phi_1=F_1(r)\e^{3i\theta}$. As a result, the
interactions Eq.~\eq{VPhiF} provide {\em three} left--handed
(right--handed) zero modes
for each of the fermions $Q$, $L$ ($U$, $D$, $E$). All of these modes are
localized in the core of the vortex. These modes were discussed in
detail in Ref.~\cite{LT}; up to the four--dimensional plane waves they
are: 
$$
Q\sim
\sum_a
\left(
\begin{array}{c}
0\\
q_{3-a}(r){\rm e}^{i(3-a)\theta}\\
q_{a-1}(r){\rm e}^{-i(a-1)\theta}\\
0
\end{array}
\right), 
~~~~
D\sim
\sum_a
\l
\begin{array}{c}
d_{a-1}(r)\e^{-i(a-1)\theta} \\
0\\
0\\
d_{3-a}(r)\e^{i(3-a)\theta} \\
\end{array}
\r,
$$
where $a=1,2,3$ enumerates three modes (three fermionic generations), and each
element of the spinor corresponds to a column of two elements,
so that the spinors are eight--component. Modes for other fermions
have the similar form with replacement $Q\to L$, $q\to l$; $D\to U$,
$d\to u$; $D\to E$, $d\to e$. The radial
functions have the following leading behaviour:
$$
f_p(r) \sim r^p, r\to 0; ~~~
f_p(r)\sim \e^{-g_f vr}, r\to\infty,
$$
where $f$ stands for $q$, $u$, $d$, $l$, and $e$.

{\bf 3.}  Masses of the four--dimensional fermions are generated by
means of the interactions \eq{V2}, \eq{V3}. As we will see
immediately, the terms in $V_2$, Eq.~\eq{V2}, are responsible for the
diagonal mass terms, while $V_3$, Eq.~\eq{V3}, generates mixings.  The
field $X$ carries exactly the same gauge quantum numbers as $\Phi$ and
is necessary to keep terms in both $V_2$ and $V_3$ gauge invariant.

To obtain the effective four-dimensional mass matrix, one has to
perform the integration over extra dimensions. Let us discuss first
the mass matrix of the down type quarks ($d$, $s$, $b$).  From the
corresponding terms in Eq.~\eq{V2} one gets the mass matrix elements
$$
m^{d(1)}_{ab}= 
Y_d\!\int\!\! r\, dr\,d\theta\,H(r)X(r)q_{3-a}(r) d_{3-b}(r)
\e^{i(a-b)\theta}=m^d_{aa}\delta_{ab};
$$
Eq.~\eq{V3} produces the contribution
$$
m^{d(2)}_{ab}=
Y_d\epsilon_d\!\int\!\! r\, dr\,d\theta\,H(r)F(r)q_{3-a}(r) d_{3-b}(r)
\e^{i(1+a-b)\theta}=m^d_{a,a+1}\delta_{a+1,b}.
$$
We see that the selection rules coming from $\theta$--dependence of
the way functions restrict the mass matrix to the form
\[
m^d=
\left(
\begin{array}{ccc}
 m_{11}^d&m_{12}^d&0\\
0&m_{22}^d&m_{23}^d\\
0&
0&m_{33}^d
\end{array}
\right)\;.
\] 
As was discussed in Ref.~\cite{LT}, the hierarchical
pattern of masses is reproduced when the profile of the Higgs field
$H$ in extra dimensions is narrow ($h\gg
g_f$, $h\gg\sqrt{\lambda}$). In this case, one can estimate, to the
leading order, the elements of the mass matrix as
\[
m^d_{aa}=2\pi Y_d\! \int\!\! r dr H(r) X(r) q_{3-a}(r)d_{3-a}(r)
\sim(\sigma_q\sigma_d)^{(3-a)},
\]
\[
m^d_{a,a+1}=2\pi Y_d\epsilon_d\! \int\!\! r dr H(r) F(r) q_{3-a}(r)d_{2-a}(r)
\sim\bar\lambda\sigma_q^{3-a}\sigma_d^{2-a},
\]
in terms of small parameters $\sigma_d=g_d/h$,
$\bar\lambda=\sqrt{\lambda}/h$. Hence,
\begin{equation}
m^d\sim Y_d\left(
\begin{array}{ccc}
\sigma_q^2\sigma_d^2&\epsilon_d \sigma_q^2\sigma_d\bar\lambda&0\\
0&\sigma_q\sigma_d&\epsilon_d \bar\lambda\sigma_q\\
0&0& 1
\end{array}
\right).
\label{m*}
\end{equation}
In a similar way, one obtains the mass matrices for up--type quarks
($u$, $c$, $t$) and charged leptons ($e$, $\mu$, $\tau$),
\begin{equation}
m^u\sim Y_u\left(
\begin{array}{ccc}
\sigma_q^2\sigma_u^2&0&0\\
\epsilon_u \sigma_q\sigma_u^2\bar\lambda&\sigma_q\sigma_u&0\\
0&\epsilon_u \bar\lambda\sigma_u& 1
\end{array}
\right), ~~~
m^l\sim Y_l\left(
\begin{array}{ccc}
\sigma_l^2\sigma_e^2&\epsilon_l \sigma_l^2\sigma_e\bar\lambda&0\\
0&\sigma_l\sigma_e&
\epsilon_l \bar\lambda\sigma_l\\
0&0& 1
\end{array}
\right).
\label{m**}
\end{equation}

These mass matrices correspond to Cabibbo-Kobayashi-Maskawa (CKM) mixing
matrix of the form:
\begin{equation}
U^{CKM}\!\simeq\!
\left(
\begin{array}{ccc}
1-{1\over 2} \zd^2&\zd-\emi\zu&-\zd\zu\emi\\
\ep\zu-\zd&1-\zd^2+2\cos\phi\zd\zu&\zd-\emi\zu\\
\zd^{2}-\ep\zu\zd&\zu-\ep\zd&1-{1\over 2}\zd^2+\ep\zd\zu
\end{array}
\right),
\label{ckm}
\end{equation}
where 
$\zd\sim|\epsilon_d|\bar\lambda\sigma_q$,
$\zu\sim|\epsilon_u|\bar\lambda\sigma_q\sigma_u^2$, and
$\phi=\arg\epsilon_u$ (all but one complex phases of the parameters
are unphysical)\footnote{Note that $\zd$ and $\zu$ are of different
order in $\sigma$; this reflects the fact that the symmetry of mixing
matrices is not $u_L\leftrightarrow d_L$, but $u_L\leftrightarrow
d_R$.}. 

As follows from Eqs.~\eq{m*}, \eq{m**}, and \eq{ckm}, to the leading
approximations, masses of nine charged fermions of the Standard Model
are determined by six parameters: the overall mass scale, say, $Y_dv_H
v_X$; two ratios of six--dimensional Yukawa couplings, $Y_u/Y_d$ and
$Y_l/Y_d$; and three combinations $(\sigma_q \sigma_u)$, $(\sigma_q \sigma_d)$,
$(\sigma_l \sigma_e)$. Four independent parameters of CKM matrix are
determined by $\zu$, $\zd$, and $\phi$.

{\bf 4.}
To conclude, we presented a mechanism to obtain three fermionic
generations together with their hierarchical masses and mixing angles,
from a single generation in a higher--dimensional theory. This
mechanism is developed along the lines of Ref.~\cite{LT} but does not
require explicit breaking of the symmetry which gives rise to
nontrivial topology of the brane. Due to higher--order interactions
between scalar field and fermions, three zero modes are localized on a
defect with topological number one.

In our particular model, the fermions are localized on an
Abrikosov--Nielsen--Olesen vortex.  With obvious modifications, the
mechanism works for other topological defects, such as a global
vortex, a monopole, or a hedgehod.  The mechanism presented above can
be embedded either in a theory with large extra dimensions
\cite{Dimop} or in a model where gravity is localized on a topological
defect \cite{loc-grav}.

M.L.\ and S.T.\ thank Universit\'{e} Libre de Bruxelles for kind
hospitality. This work was partially supported by the ``Actions de Recherche
Concret\'ees'' of ``Communaut\'e Fran\c{c}aise de Belgique'' and
IISN--Belgium.  The work of M.L.\ and S.T.\ is supported in part by
RFFI grant 99-02-18410a, by CRDF award RP1-2103, by the Russian
Academy of Sciences, JRP grant No.~37, by the Council for
Presidential Grants and State Support of Leading Scientific Schools, grant
00-15-96626, and by the programme SCOPES of
the Swiss National Science Foundation, project No. 7SUPJ062239,
financed by Federal Department of Foreign affairs.  The work of S.T.\
is supported in part by Swiss Science Foundation, grant 21-58947.99.

\end{document}